# NARRATING NETWORKS

## Exploring the Affordances of Networks as Storytelling Devices in Journalism

**Bounegru, Liliana, Tommaso Venturini, Jonathan Gray, and Mathieu Jacomy**

*Networks have become the de facto diagram of the Big Data age (try searching Google Images for [big data AND visualisation] and see). The concept of networks has become central to many fields of human inquiry and is said to revolutionise everything from medicine to markets to military intelligence. While the mathematical and analytical capabilities of networks have been extensively studied over the years, in this article we argue that the storytelling affordances of networks have been comparatively neglected. In order to address this we use multimodal analysis to examine the stories that networks evoke in a series of journalism articles. We develop a protocol by means of which narrative meanings can be construed from network imagery and the context in which it is embedded, and discuss five different kinds of narrative readings of networks, illustrated with analyses of examples from journalism. Finally, to support further research in this area we discuss methodological issues that we encountered and suggest directions for future study to advance and broaden research around this defining aspect of visual culture after the digital turn.*

## Introduction

As legend has it, a few days after the attacks of September 11, 2001 in New York, an FBI agent showed up at the Whitney Museum of American Art, and asked to examine a painting of the conceptual artist Mark Lombardi. Before committing suicide in March 2000, Lombardi painted a series of canvases portraying the networks of financial and political power around fraud or corruption events reported in the news. According to the legend, the FBI agent was particularly interested in the painting "George W. Bush, Harken Energy, and Jackson Stephens, ca 1979–90," showing the multiple relations connecting the former US president and the terrorist sheikh Osama bin Laden.

Whether true or not, this legend says something of the role that networks have come to occupy in our societies. Over the past couple of decades, networks, understood as sets of nodes (or vertices) connected by edges (or links), have become an increasingly popular way to represent all kinds of collective phenomena. They have become the *sine*

*qua non* of visual culture in the digital age, illustrating and attesting to the complex webs of association around us. The metaphors and imagery of networks have exploded across many fields and are said to have the potential to revolutionise everything from medicine to markets to military intelligence. The ubiquity of the Internet as a network of computer networks, of the Web as a network of hypertexts, of relational databases and of the rise of social media platforms facilitating mass connectivity (as well as its traceability) have all contributed to this renaissance of networks (Barabási 2002; Watts 2004; Rieder 2012; Munster 2013).

Following Espeland's (2015) call to scrutinise not just the analytical operations that make up quantification practices but also the circulation, interpretation, impact and narratives that numbers generate, in this article we look beyond the mathematical properties of networks (which have received extensive attention for centuries – see, e.g., graph theory and sociometry) towards the stories that they evoke. To do so, we study the narratives constructed around network visualisations, hereafter called networks, in a series of journalistic projects and articles.

We focus on journalism because networks have been capturing the journalistic imagination for a few years now. As journalism is increasingly being reconfigured as a data public (Ruppert 2015), called to organise its practices around the collection, analysis and narrativisation of data, the analytical and communicative potential of networks has received growing journalistic attention. We also focus on journalism because it has long been associated with practices of storytelling: the craft of rendering complex phenomena into narrative form (for an overview see Kormelink and Costera Meijer (2015); Zelizer (2004)). In a 1982 article, Michael Schudson argues that the power of media stems not from the delivery of "facts," but rather from the development of narratives and narrative forms, i.e. conventions by means of which sequences of actions or events are being reconstructed, ordered and presented. "Journalists are professional storytellers of our age," writes Bell (2005, 397). Journalistic storytelling has been conceived and studied in multiple ways. One strand of studies focuses on narrative styles, forms and conventions by means of which journalists decide what becomes news and how it is structured (see, e.g., Tuchman (1976); Darnton (1975); Bell (2007)). A famous such style is the "inverted pyramid" (Bell 2007). Others focus on the myth-making capacities of journalism, i.e. its capacity to create representations of the world that produce or reproduce ideological or value systems and are generative of character types such as the hero and the villain (see, e.g., Knight and Dean (1982); Ettema and Glasser (1988)). In recent years the digitisation of journalism has been associated with renewed possibilities for narration, emphasising possibilities for more immersive, interactive, multimodal and participatory forms of storytelling (Pavlik 2000, 2001).

**When Networks Meet Narratives**

While the use of networks and network visualisations for data exploration and analysis has been studied extensively (see, e.g. Adamic and Adar 2003, Andris et al. 2015), the narrative or storytelling potential of networks is just beginning to receive more sustained attention from researchers.

Networks and narratives have recently come together in a number of different areas of research. One such area is that of information and communication technology and organisation studies, where narrative networks represent methodological devices for representing patterns and routines that emerge around the usage of information technologies (Pentland and Feldman 2007; Weeks 2012). Another perhaps less expected direction of study is exemplified by the Narrative Networks programme set up by U.S. government's Defense Advanced Research Projects Agency (DARPA). The programme aims to explore the relationships between narratives, human cognition and behaviour in the context of international security. Another more familiar area of inquiry that brings together these two concepts is narrative theory. In this context network analysis methods have been applied to the study of literary texts in attempts to develop quantitative approaches to enrich the study of narrative texts (see, e.g., Bearman and Stovel 2000; Moretti 2011; Sudhahar et al. 2015).

Our interest differs from these other approaches in that we are predominantly interested in the stories that networks may elicit. This is an area of research that is just beginning to be explored (see, e.g., [name deleted to maintain the integrity of the review process]; Bach et al. 2016; Suslik Spritzer 2015). To distinguish our approach from the various strands of work developed around the concept of narrative networks, we propose the complementary notions of *network narratives* and *network stories*. While distinctions between "narrative" and "story" have been drawn in narrative theory (for an explanation see, e.g., Culler 2001), in this article we use the terms "stories" and "narratives" interchangeably. The notion of network narrative or network story is intended to guide attention towards the ways in which networks may be "storified," i.e. towards the ways in which narrative meaning may be constructed around them. By proposing this notion we do not aim to suggest that network visualisations *are narratives*. Instead our aim is to draw attention, following Ryan (2004), to their potential to *possess narrativity*. Ryan characterises the distinction as follows:

> The property of 'being' a narrative can be predicated on any semiotic object produced with the intent of evoking a narrative script in the mind of the audience. 'Having narrativity,' on the other hand, means being able to evoke such a script. In addition to life itself, pictures, music, or dance can have narrativity without being narratives in a literal sense (2004, 9).

**An Analysis of Journalistic Network Stories**

We started our analysis by building a collection of journalism pieces which use network diagrams and concepts.[1] We collected 45 exemplary journalism pieces from a number of different sources: interviews with journalists working in this area, online repositories of journalism stories such as those associated with the Data Journalism Awards competition[2] and with the National Institute for Computer-Assisted Reporting (NICAR),[3] and specialty mailing lists such as NICAR-L[4] and Influence Mapping.[5] These pieces belong to various journalistic genres, from visual and investigative journalism, to special reports and interactives, to name just a few, and are published both in print and digital formats. In these pieces network graphics are integrated in multimedia packages or used to illustrate pieces of writing. The stories and network graphics which we collected were not always accessed in their original medium of publication and hence some of the elements of context in which they were originally embedded are absent. However, since these elements of context are essential to the interpretation of the narrative, we only considered in our analysis those network graphics for which sufficient elements of context were available. Given the illustrative nature of our analysis we ended up focusing on 13 of the 45 pieces that were identified in the collection that we created.

---

[1] For the purposes of this article we use the terms "network diagram," "network visualisation" and "network graphic" interchangeably.
[2] Accessible at: http://www.globaleditorsnetwork.org/programmes/data-journalism-awards/
[3] Accessible at: https://www.ire.org/nicar/
[4] Accessible at: https://www.ire.org/resource-center/listservs/subscribe-nicar-l/
[5] Accessible at: https://groups.google.com/forum/#!forum/influencemapping

While not the primary focus of our research, we drew on some of the resources of multimodal analysis (Jewitt 2014; Kress and Van Leeuwen 2001) to construe narrative meaning. Multimodality refers to "approaches that understand communication and representation to be more than about language, and which attend to the full range of communicational forms people use – image, gesture, gaze, posture and so on – and the relationships between them" (Jewitt 2009, 14). The network stories that we selected are realised through the interaction of multiple modes, from static and interactive diagrams, to photographs, pictograms, and written language, the latter being present in various forms, such as story headline, lead, body, graphic caption, labels, legend and instructions. All these modes are meaningfully organised in the layout space of the journalistic piece. A mode is defined as "a socially shaped and culturally given resource for making meaning," utilized in representation and communication (Kress 2014, 60).[6]

While multimodal analysis is a well-documented area of research (see, e.g., Jewitt 2014), in what follows we would like to propose a model of analysis for the graphic representation of networks, an area which is only beginning to be addressed in multimodal research. To construe narrative meaning from network graphics, we draw on Bertin's semiotic model for the analysis of graphics (1983), and on [name deleted to maintain the integrity of the review process]'s framework for the visual analysis of networks (2015). Bertin proposes that the reading of a graph largely consists of constructing correspondences around a central notion called the invariant, characterised by means of a series of visual variables (such as size, shape, colour and texture) and that it occurs in stages (1983, 140-141). Crucial to the creation of such correspondences and thus to the construal of narrative meaning is a *question*, conscious or not, which guides the reader's perception towards the relevant associations. [Name deleted to maintain the integrity of the review process] apply Bertin's model to network graphics and propose a framework for constructing meaning around three visual variables of network graphs: node position, node size and node hue. In the next section we illustrate how we applied these models to the analysis of multimodal journalistic stories with a network graphical component.

Another essential device in narrative construction and interpretation in our corpus is the genre of journalistic storytelling. Following the convention of this genre that the key points are presented at the beginning of the story, particularly in the headline and the lead paragraph (Bell 2007), where applicable, we have taken guidance in the formulation of the question that the network graphic invites the reader to ask, and hence of the story which it evokes, from the headline and lead of the journalistic piece, as well as from the graph elements addressed in textual form in the body of the story. Finally, the socio-cultural knowledge of the reader also plays an important role.

Drawing on these models, we take a number of steps in our analysis. Given our interest to explore the stories that networks may elicit in journalism pieces, the starting point of our analysis is to identify the notions, themes or concepts central to the journalism piece that are expressed through the network graphic, or what in multimodal research may be termed the diagrammatic mode (Alshwaikh 2009). Next we detect the prominent visual attributes of the network visualisation which guide readers' perceptions and by means of which narrative meaning is cued. The narrative view is then further specified and qualified by means of the textual elements in close proximity to the graph, most importantly its caption and the title, lead paragraph and body of the journalistic piece in which it is embedded, also mobilising more broadly our knowledge of the world and of the journalistic genre. We verbally formulate the narrative views which we construe and where applicable draw associations between them and the network concepts or properties which they deploy or with which they resonate.

---

[6] For a more extensive discussion of semiotic modes see, e.g., Hiippala 2014, Kress 2014 and Bateman 2011.

To move from narrative readings to narrative reading types we use an emergent categorisation approach whereby categories of narrative readings are being construed by identifying repetitive story types or patterns in our collection of journalism pieces. Given the qualitative nature of our study focused on demonstrating the narrative potential of networks, we do not exhaustively analyse our collection of journalism pieces. Instead we focus on the construal of an illustrative set of narrative reading types, each of which is illustrated with the analysis of three examples.

**Five Ways Networks are Used for Journalistic Storytelling**

Our qualitative analysis shows that there are recurring narrative reading types that networks elicit and that these narrative readings resonate with or deploy a number of network properties or concepts. In this section we discuss five such examples of narrative readings and illustrate them with the analysis of three cases where they occur. As the analysis below shows, multiple narrative readings may be construed from a journalistic piece, which means that a journalism project or article may be discussed in relation to multiple categories. The construal of multiple narrative views occurs particularly in the case of pieces which deploy a mode which could be tentatively termed "dynamic diagram." This composite mode incorporates a strong interactive component which helps to connect different narrative sequences and enables the narrative to progress. Unlike a static diagram, the dynamic network diagram enables the user to explore and manipulate the graph display through a variety of interactive devices and techniques. Given the illustrative nature of our proposed typology of narrative readings, we do not undertake to exhaust all of the narrative readings of a journalistic project. Instead we focus on the ones that best illustrate the narrative views under examination.

*1. Exploring Associations around Single Actors*

In this category we grouped narrative views that depict the network of entities around a single actor. This category may be interpreted as evoking a particular type of network graph, often called "ego-network." One typical characteristic of ego-networks is the depiction of relationships around a given social unit, referred to as the *ego*, resulting in "a mini-network or immediate neighbourhood surrounding the ego" (Freeman 1982, 291).

We also distinguish this category from the second narrative view which depicts key players, in that the nodes that play the role of the ego in this category of stories are not necessarily well-connected ones (authorities or hubs), which is where the emphasis lies in the "detecting key players" category in the next section. In borderline cases, where it has been difficult to place a story in one category or another, we returned to the written language mode and particularly to the journalism piece headline and lead as well as to the graphic headline to identify additional cues for the construal of the narrative view.

We will discuss three cases of narrative views which we have construed as "exploring associations around single actors" in our collection of journalism pieces. Across all these cases, crucial to the construal of the selected narrative reading is the interactive component of the dynamic diagrammatic mode. Interactive techniques such as changing the appearance of the arrow cursor into a hand (known as the pointer cursor), mouse-over to reveal details-on-demand, implicit instruction through visualisation guides and explicit instruction through mouse-hover over elements of the graph, guide the reader to select single nodes in order to explore their networks of associations.

*Thomson Reuters*' "Connected China" project (February 2013)[7] is an interactive multimedia website dedicated to "tracking thousands of people, institutions and connections that form China's elite power structure" by depicting networks of familial, political and social ties amongst members of its governing structures. Of the multiple narrative readings elicited by the dynamic diagrammatic mode, in this section we describe a prominent one, the exploration of associations around single actors. This narrative reading draws on multiple modes. In the written language and layout modes the importance of the ego is expressed through elements such as the guide to the reading of the visualisation (see figure 1(a)), the textual explainers describing each individual in the network and the labels identifying them by name. In the diagrammatic mode, an important cue is the prominent positioning of the ego in the graph, its representation through a photograph of the individual whom it represents, and the disposition of the related nodes in a semicircle around it (see figure 1(b)).

[Figures 1(a) and 1(b) near here]

In the interactive micro-site "WESD Web of Connections"[8] published by the *Statesman Journal* (n.d.), the Willamette Educational Service District's network of connections is a multi-layered story about corruption, oversight and abuse undertaken by a regional education service agency in the United States. Multiple modes participate in the construction of this narrative reading. In the default view of the interactive network graphic, the Willamette Educational Service District is set as the ego but users can move any other node to the centre of the graph by clicking on it. The written language mode, more specifically the graph title, instructions and labels, interacts with the dynamic diagrammatic mode and its spatial disposition of nodes and edges to guide the interpretation of the network around a single actor (see figures 2(a) and (b)). The primary visual property through which the actor's network is realised is the spatial disposition of the ego at the centre of a radial graph, with edges radiating from it and connecting it to nodes representing suspect deals, administrative structures and the individuals, organisations and addresses associated with these.

[Figures 2(a) and 2(b) near here]

Finally, *Washington Post*'s "Top Secret America,"[9] an investigative project published on July 19, 2010, enables the user to explore the network of types of work conducted by and companies contracted by each of the 45 government organisations that make up the national security program of the United States. This narrative reading is cued by the written language and dynamic diagrammatic modes. An interactive device, an information box appearing upon hovering over elements of the visualization, guides the reader to click on the name of an agency in order to explore its work areas and contractors (see figure 3(a)). The ego is cued in the diagrammatic mode by its spatial disposition at the centre of the graph, as well as the size of the node representing it. Its network of entities is depicted through a series of circles of varying sizes and blocks of varying colours. The ego and its network of elements are further qualified in the written language mode through the graphic's caption, the information box and the node label (see figure 3(b)).

---

[7] Accessible at: http://china.fathom.info/
[8] Accessible at: http://community.statesmanjournal.com/news/wesd/web/
[9] Accessible at: http://projects.washingtonpost.com/top-secret-america/

[Figures 3(a) and 3(b) near here]

*2. Detecting Key Players*

In this category we grouped narrative views that depicted key actors based on the number of connections with other nodes. The focus of these network stories is on the density of connections around one or several central nodes. This category deploys the network property of power law distribution (Pareto 1965; Barabási 2002). The Pareto or power law distribution of connectivity indicates the concentration of a large majority of connections around a small minority of nodes. Such nodes are called authorities or hubs, to show that they receive or spawn an unusually large number of links. Such a distribution of associations can be observed in the topology of the web, where a few webpages receive millions of incoming links, whereas close to 90% of all webpages receive ten or less incoming links (Barabási 2002). This is also identified as a property of many other types of self-organizing networks in biology, economy and society.

We will illustrate this narrative view with three examples. The first one is the *New Scientist*'s "The Stem Cell Wars" special report[10] on citation practices in stem cell science, published on June 12, 2010. In this example the written language mode contributes to the narrative reading in multiple ways. The graphic caption guides the reading of the graphic around the detection of "influential players" (see figure 4(a)). The information box "The strongest link" also cues this reading in its first sentence: "Shinya Yamanaka of Kyoto University in Japan is the dominant scientist in cellular reprogramming" (*New Scientist*, June 12, 2010; see figure 4(b)). The most cited scientists, the Japanese Yamanaka and the U.S. based Jaenisch and Hochedlinger, are also discussed extensively in the body of the article. In the static network diagram, the node size as well as the depiction of edges as arrows pointing towards the most cited scientists gestures towards this narrative reading. The edges here represent citations received by papers authored by the scientists depicted as nodes in the graph.

[Figures 4(a) and 4(b) near here]

The article series entitled "Park Young-Joon at the Center of President Lee Myung-bak's Human Resources Network," published on January 20, 2002, is a *JoongAng Ilbo* investigation into the social network of senior officials around South Korea's president, Lee Myung-bak. In the written language mode, the headline of the article,[11] the caption

---

[10] Accessible at: http://www.peteraldhous.com/Articles/The_stem_cell_wars.pdf
[11] In translation: "In this social network among the 944 senior officials, Park Young-joon is located at the center. An official is positioned toward the center when he or she has

of the graphic,[12] and the body of the article all point towards the strength of associations of government official Park Young-joon to other top officials with links to the president. These links include whether they share the president's hometown, whether they graduated from the same university, whether they served on the president's campaign and a number of other business and political ties. In the static diagram the position of the node representing the most connected government official at the centre of the graph and the size of the icon representing him supports such a narrative reading (see figure 5).

[Figure 5 near here]

Our final example in this section is the "Flip Investigation"[13] published by the *Sarasota Herald Tribune* in 2009, examining real-estate fraud in the state of Florida. While the dynamic diagrammatic mode invites multiple readings, in this section we focus on the "detecting key players" narrative view, which is prominently cued in the article series accompanying the interactive network visualisation. Here the key player is construed as a villain situated at the centre of a network of fraudulent transactions known as property flips. The narrative reading is signalled in the written language and photograph modes, with a full text article in the series, published on July 21, carrying the evocative title of "The King of the Sarasota Flip," being dedicated to the investigation of the key player, Craig Adams (see figure 6(a)).[14] In the diagrammatic mode, the narrative reading is realised through the default frame of the "Network" view of the interactive graph depicting Craig Adams at the centre of the flip deals network, comprising of other real-estate professionals, "flippers," orchestrators as well as victims (see figure 6(b)). The topology of the network comes very close to the star-shaped "winner takes all" topology (Barabási 2002), in which the central hub is connected to all the other nodes.

[Figures 6(a) and 6(b) near here]

*3. Mapping Alliances and Oppositions*

In this category we grouped narrative views that depict associations of nodes as well as the absence of associations between groupings of nodes. This category deploys the network property of clustering, which measures the varying density of connections between nodes (Watts and Strogatz 1998) as well as the property of structural holes (Burt 1992). A cluster is thus a collection of nodes that are more densely connected among each other than to the rest of the network. A cluster is visually displayed by the spatial disposition of such nodes in proximity to each other in a network graph. According to Burt the concept of structural hole refers to the "separation between nonredundant contacts" where nonredundant contacts are nodes that do not share connections (1992, 18). In the visual representation of a network, the clusters may be construed as alliances between actors and the absence of connections or structural holes

---

more links to President Lee." Source: NICAR Stories Database, story no. 25691.
[12] In translation: "In this social network among the 944 senior officials, Park Young-joon is located at the center. An official is positioned toward the center when he or she has more links to President Lee." Source: NICAR Stories Database, story no. 25691.
[13] Accessible at: http://projects.heraldtribune.com/investigateflip/investigateflip.html
[14] Accessible at: http://www.heraldtribune.com/article/20090721/ARTICLE/907211055

may be construed as opposition or lack of allegiance ([name deleted to maintain the integrity of the review process] 2015).

In what follows we will illustrate the construal of this narrative view with three examples from the political domain. *Le Monde*'s piece "2007-2011: La Cartographie de la Blogosphère Politique,"[15] is an interactive graphic published on July 4, 2011 which represents the linkages amongst the French political blogosphere between 2007 and 2011. In the diagrammatic mode, the visual property through which the reading of alliances or coalitions between actors is realised is the spatial disposition of nodes in the graph into clusters. The clusters are identified through the density of associations amongst nodes, and through their colour (see figure 7). Alliances may also be read not only between nodes but also between clusters. Such alliances can be read from the position of clusters in relation to one another. Clusters that are closer and share more links (such as the greens and the left bloggers' clusters in the 2011 map) can be read as allies whereas the absence of links or the presence of structural holes between two clusters cues the reading of opposition (as in the case of the sparse connections between the extreme right on the one hand, and greens and left bloggers on the other in the 2011 map).

The journalistic genre convention that the most important information is presented at the beginning of the piece provides essential guidance, as the network of clusters is the default view of the graph. The alliances and oppositions are further qualified through the written language mode. The title or headline of the piece identifies the graph as a map of the French political blogosphere. An interactive component, the navigation menu on the left-hand side of the graph identifies the political factions textually through their labels ("gauche," "extreme gauche," "ecologie politique," "centre," "droite," and "extreme droite")[16] and pictorially through the colour of the icon associated with each political faction, which in turn corresponds to the colour of the clusters on the map. The reading guide which accompanies the piece provides further guidance on the interpretation of the position of the nodes in relation to each other:

> Ce type de placement permet de rendre visibles les dynamiques communautaires et le fait que certains sites échangent fortement entre eux et beaucoup moins avec les autres sites de la carte (qui seront donc plus éloignés)" (*Le Monde*, 4 July, 2011).[17]

[Figure 7 near here]

*Global News*' piece "Visualizing the Split on Toronto City Council," published on March 20, 2012,[18] tells the story of the growing divergence between Toronto council members between 2010 and 2012. Similarly to the previous example, the alliances and oppositions are rendered in the diagrammatic mode through the spatial disposition of the nodes in separate clusters. The clusters are represented through different colours (red and blue) and are bridged by a few nodes in purple (see figure 8). The narrative reading is further qualified as representing the alliances and oppositions around voting on the Toronto City Council through the written language mode, more specifically the headline of the article ("Visualizing the Split on Toronto City Council"), and the body of the article which identifies the factions as follows: "blue for the group around the mayor, red for the opposition, and purple for centrist or unaffiliated councillors" (*Global News*, 20 March, 2012). The principle of association and opposition is also identified in the written language mode: "Councillors who tend to vote together will be clustered closely together in the graphic" (*Global News*, 20 March, 2012). The coalitions and oppositions are calculated and represented at different moments in time. This constitutes another narrative view type, which we discuss in our fourth category below.

---

[15] In translation: "2007-2011: The Cartography of the Political Blogosphere"; accessible at: http://www.lemonde.fr/election-presidentielle-2012/visuel/2011/07/04/la-cartographie-de-la-blogosphere-politique_1544714_1471069.html

[16] In translation "left," "extreme left," "greens," "right" and "extreme right" respectively.

[17] In translation: "This type of disposition helps to make visible community dynamics and the fact that some sites exchange more links among themselves than with the other sites on the map (which will therefore be more distant)"

[18] Archived version accessible at: https://web.archive.org/web/20120507224159/http://www.globalnews.ca/topics/torontocouncildiagrams/dec_2010/index.html

[Figure 8 near here]

Lastly, the *Le Monde* piece, "Mariage Gay: L'Opposition Soigne ses Amendements," published on January 31, 2013,[19] investigates coalitions formed around amendments to the gay marriage law proposal in France. The diagrammatic mode signals the reading of coalitions through the spatial disposition of the nodes in several clusters in the interactive network graph which concludes the article. Political parties are represented through different colours and largely correspond to the clusters representing coalitions around the signing of legislative amendments (see figure 9). The reading of coalitions is further specified in the headline and body of the article, as representing the alliances around the signing of amendments to the gay marriage law. The alliances are shown to follow ideological lines, with co-signing occurring largely within the boundaries of the political party, with the exception of the right-wing UMP (Union for a Popular Movement), which co-signs with independent parliament members as well as with the centrist UDI (Union of Democrats and Independents). The absence of connections between centre-left parties such as the RRDP (Radical, Republican, Democratic and Progressist) and ECOLO (Ecologists), as well as between centre-left and right-wing in the diagram can be construed as opposition. The principle of association between nodes in the graphic is also identified in the written language mode as follows:

> Chaque cercle représente un parlementaire. Lorsqu'un député a signé ou cosigné un amendement avec un autre, un lien les relie. Les cercles sont ensuite placés de telle sorte que les cercles attachés les uns aux autres se rapprochent et que ceux qui ne sont pas attachés s'éloignent. La taille du cercle varie en fonction du nombre d'amendements signés par le député. (*Le Monde*, January 31, 2013)[20]

---

[19] Accessible at: http://www.lemonde.fr/societe/article/2013/01/31/mariage-pour-tous-l-opposition-soigne-ses-amendements_1825467_3224.html. In translation the title reads: "Gay Marriage: The Opposition Signs its Amendments."

[20] In translation: "Each circle represents a member of the Parliament. When a member has signed or co-signed an amendment with another, a link between them is established. The circles are then placed

[Figure 9 near here]

*4. Exploring the Evolution of Associations over Time*

In this category we grouped narrative views formed around a temporal dimension and which show the transformation of associations of actors over time. This category deploys a property that is common to real-world networks, namely that they are dynamic systems whose composition and topology are subject to change over time (Barabási 2002). In what follows we will illustrate this narrative view with three examples, two of which have been encountered in previous categories as well. In the case of all three examples, essential to the construal of this narrative reading are the composite modes of the dynamic diagram or the animated diagram (as is the case of the third example) with their interactive component. Interactivity is realised through devices such as a navigation menu indicating different moments in time in the case of the first two examples and a time bar or progress bar in the case of the final example, all of which enable the reader to navigate across visualisation panels.

In *Le Monde*'s "2007-2011: La Cartographie de la Blogosphère Politique," (July 4, 2011) we identify a second narrative view, which we name "exploring the evolution of associations over time." The title or headline of the piece identifies the time period covered by the mapping: 2007 to 2011. An interactive device, the navigation menu (containing three buttons: "2011," "2009," and "2007") and the caption of each visualisation panel ("le web politique militant en 2007," 2009 and 2011 respectively) specify the particular year represented by the panel. The narrative view is also cued pictorially by the representation of network dynamics at two-year intervals through three maps, which the user can navigate from a menu at the top of the interactive graphic (see figures 10(a), 10(b) and 10(c)).

[Figures 10(a), 10(b) and 10(c) near here]

The narrative view "exploring the evolution of associations over time" is similarly construed in the piece "Visualizing the Split on Toronto City Council" (*Global News*, March 20, 2012). One important interactive element which cues this narrative reading is the navigation menu which enables the user to switch between representations of the network at various moments in time (December 2010, May 2011, January 2012, February 2012, March-April 2012), as well as the caption of each graph, which indicates the moment in time which it depicts (see figures 11(a), 11(b) and 11(c)). The longitudinal or temporal dimension is further qualified in the body of the article, where the key shifts in network dynamics at different moments in time are being analysed.

[Figures 11(a), 11(b) and 11(c) near here]

Lastly, in the *Guardian*'s interactive piece "How Riot Rumours Spread on Twitter" (December 7, 2011),[21] an investigation into the emergence and correction of misinformation pertaining to the 2011 UK riots on Twitter, this narrative view is realised primarily with the participation of the dynamic diagrammatic mode, the written language mode and the layout mode. An interactive device, the time bar or progress bar

---

so that the circles linked to each other are closer. The size of the circle is proportional to the number of amendments signed by the member of Parliament."

[21] Accessible at: http://www.theguardian.com/uk/interactive/2011/dec/07/london-riots-twitter

present on the web page dedicated to each rumour, enables the reader to follow the development of the rumour from its inception until its death. The diagrammatic mode signals the similarity of tweets through their spatial proximity in the graph, as well as their position in relation to the rumour (support, opposition, questioning and commentary), through the property of colour (see figures 12(a), 12(b) and 12(c)). This narrative reading is further qualified through textual annotation highlighting key events in the unfolding of the rumour as well as through the intensity of the nodes' colour, whereby lighter tones are used to represent more recent tweets and darker tones are used to represent older tweets.

[Figures 12(a), 12(b) and 12(c) near here]

## *5. Revealing Hidden Ties*

In this category we grouped narrative views which depict hidden and potentially incriminating sequences of connections or paths between nodes. In this category the nodes represent (alleged) villains as well as their collaborators, whether individuals, businesses or government officials. The edges represent allegedly unethical, fraudulent and potentially incriminating relations. The painting by Marc Lombardi cited in the introduction clearly falls in this category.

This category may be interpreted as loosely evoking the network property of weak ties. Whereas the nodes of a cluster are connected by strong ties, another essential concept in the study of social networks is that of weak ties (Granovetter 1977). The notion is used to describe connections between nodes belonging to different clusters where nodes through which such connections are established act as bridges. Such nodes have been shown to play an essential role in various social activities through their ability to transport information across the structural holes that separate clusters (Jensen et al. 2015). This narrative view should be distinguished from "exploring associations around single actors" and "detecting key players" in that, while selected actors are at the centre of such narrative views, the emphasis of the reading is not on one or several well-connected actors, but rather on the path of connections that ties the actors and the nature of these ties. We will illustrate this narrative view below with three examples.

*The Kansas City Star*'s "Terrorist Tentacles Know no Boundaries" (November 28, 2004) explores the ties between a global charity and multiple terrorist organisations and supporters, including terrorist Osama bin Laden. In the diagrammatic mode ties or connections are rendered through directed edges represented as arrows, all of which start from a single prominently sized node and point towards several other nodes, which can be construed as individuals or groups based on their icon depictions (see figure 13). The nodes are further qualified through textual labels which identify them as the incriminated global charity IARA (Islamic African Relief Agency) and the terrorist organisations, terrorism supporters and terrorists to which IARA is alleged to have ties, including Osama Bin Laden. The incriminating ties are further specified in the headline of the article ("Terrorist Tentacles Know no Boundaries") as well as in the body of the article where the ties are being listed:

> At least eight connections between IARA and Osama bin Laden, his organizations or the Taliban; Two connections to Hamas, the Palestinian terrorist organizations whose suicide bombings ravaged life in Israel; Connections to three other groups that long have been designated as terrorist organizations by federal authorities. (*Kansas City Star*, November 28, 2004)

[Figure 13 near here]

Similarly, in *Los Angeles Times*' piece "The Calderon Family's Connections" (February 21, 2014),[22] which investigates incriminating ties between a powerful Southern Californian family and organisations in its area, both the diagrammatic and the written language modes participate in the narrative reading of incriminating ties. The incriminating ties or connections are rendered through arrows connecting two sets of nodes (see figure 14). The nodes are qualified through textual labels as the members of the Calderon family and the organisations and businesses with which it has established incriminating relations. The nature of the relations represented by the edges is specified through colour and text: green arrows represent financial donations to campaigns, blue arrows represent legislative interventions or attempts thereof, and yellow arrows represent consultancy services. The headline of the piece ("The Calderon Family's Connections") further anchors this narrative reading. The subtitle of the article qualifies the ties as incriminating through the specification that the Calderon family is under investigation by the FBI. The nature of the ties is also cued by the body of the article, which provides further detail on the investigated connections between the family and various private and public actors in their region.

[Figure 14 near here]

Finally, we discuss the construal of incriminating ties in Organised Crime and Corruption Reporting Project (OCCRP)'s interactive piece "The Proxy Platform," published on November 22, 2011.[23] In this piece too the path is cued by the dynamic diagram and written languages modes. An interactive feature enables the reader to select a node and to visually highlight the sequence of paths or connections to other nodes, as well as to reveal the name labels of these nodes, while the other elements of the graph are dimmed. The sequence of paths is composed of four different types of nodes. The sequencing of the node types on a horizontal axis, inviting (but not restricting) the reading of paths/connections between nodes from left to right further invites the construal of the notion of a path connecting different kinds of nodes (see figure 15). The node types are specified through labels, from left to right, as: "proxies" (individuals running phantom companies on behalf of the real beneficiaries), "proxy companies" (companies set up to facilitate money laundering), "banks" through which transactions flow between proxy companies and beneficiaries, and "beneficiary companies" who are on the receiving end of financial transactions). The interpretation of the sequences of connections as incriminating ties would be highly improbable from the diagrammatic mode alone. It is the written language mode, more specifically the body of the article accompanying the interactive graph which describes the mechanisms of the money laundering system which spans Eastern Europe, Central and South America and Asia, that enables the identification of the sequences of paths between nodes as incriminating ties in a money laundering system.

[Figure 15 near here]

---

[22] Accessible at: http://graphics.latimes.com/calderon/
[23] Accessible at: https://www.reportingproject.net/proxy/en/

**Conclusions**

In this article we illustrated how journalistic storytelling is configured through the usage of network visualisations by examining the narratives that networks elicit in a series of journalism pieces. We showed that there are patterns in the narrative reading of networks, and discussed in detail five of these narrative view types: "exploring associations around single actors," "detecting key players," "mapping alliances and oppositions," "exploring the evolution of associations over time," and "revealing hidden ties."

More established forms of storytelling which integrate visualisations are held to convey the narrative primarily through text (Segel and Heer 2010). In contrast to this claim, this article showed that narrative views elicited in journalism pieces featuring network graphics are realised multimodally. It is in the interaction between modes such as static, dynamic or animated diagrams, written language, photographs, layout and pictograms that the specificity of this composite form of storytelling lies.

In keeping with Ryan's proposition that good narratives think with their medium (2005), that is that they take advantage of the affordances of the medium in which they are realised, we show that different narrative views tend to be cued by distinct visual attributes of the diagrammatic mode (e.g. node size and position for "exploring associations around single actors" and "detecting key players"; density of connections and node colour for "mapping alliances and oppositions," and size and arrow-like depiction of edges in the case of "revealing hidden ties"). Moreover, we find interactivity and the devices through which it is realised to be crucial to the construal of several of the narrative reading types, particularly "exploring associations around single actors" and "exploring the evolution of associations over time." In line with the same argument we show that narrative views deploy or evoke classic network concepts and properties, from ego-networks, to the power law distribution, clustering, weak ties and the dynamic character of real-life networks.

In addition to their resonance with network concepts or properties we also find these narrative readings to resonate with some of the values and themes associated with news and investigative journalism. For example, the narrative readings constructed around the depiction of actors' networks, key players and alliances and oppositions may be interpreted as resonating with news journalism values such as the focus on human interest, personalities, eliteness of news actors, and conflict (Bell 2007; Dardenne 2005). Narrative readings constructed around the exposure of incriminating ties resonate with conceptions of investigative journalism narratives to be acting as an instrument for invoking morality in the service of judging civic vice (Ettema and Glasser 1998).

However, while these narrative readings of networks resonate with some of the norms and values of the journalistic genres in which they are embedded, networks come with their own storytelling affordances. As the discussed examples have shown, the production of narrative meaning revolves around relationality or associations and their properties, from topology, to density, to dynamic character or absence. For this reason journalistic stories that we examined were often configured around the representation and exploration of structures, assemblages or collectives, be they political or power structures and fraud or corruption networks, to mention just a few.

Returning to the multimodal realisation of these narrative views, besides being cued by particular visual attributes of networks, in all the examined cases the identification and qualification of the actors, connections between them, themes, temporal dimensions and other elements of the narrative reading would have been improbable without anchoring in the written language mode as well as context. The written language and layout modes, present in our examples in the form of headlines, article bodies, graphic labels, captions and guides, are so important in the construal of narrative meaning that in our analysis we resolved to eliminate multiple examples of network graphics that had been used in the context of journalism pieces in cases where their original context of publication was unavailable to us and we were thus unable to unambiguously construe narrative meaning. The journalism genre and particularly its convention that the most important information is contained in the headline and lead of an article, and the subjectivity and socio-cultural knowledge of the reader more broadly, also played an important role in guiding the interpretation of narrative meaning. For this reason we place emphasis in our analysis on the description of the process by means of which we have *construed* each narrative reading.

### *Issues for further thought and research*

In what follows we would like to discuss a few methodological issues which help to further qualify our findings. Firstly, given our small collection of examples we make no claims of comprehensiveness or representativeness of the narrative views which we identified. While for the purposes of this article we limited ourselves to the discussion of five narrative views that exploit or resonate with network concepts or properties that have captured the popular network imagination today, we believe that other narrative readings may also be construed (for example around the concept of "spheres of influence") and invite researchers to conduct further work in this area. To facilitate such work we publish together with our article the full collection of examples of journalism pieces deploying network visualisations or concepts and encourage other researchers to expand it.[24]

Secondly, we also do not claim the boundaries between the identified narrative reading types to be clear-cut. In fact we encounter in our analysis several borderline cases where, on the basis of the diagrammatic mode alone, the narrative meaning can be equally construed along the "exploring associations around single actors," "detecting key players" and "revealing hidden ties" narrative views. In these cases we anchor or qualify the narrative reading by taking into account information present in other elements of the journalistic piece, particularly the headline, lead and body of the article.

Thirdly, we would like to note that the narrative views which we identified are not mutually exclusive. In fact in our collection we were often able to construe multiple narrative views from a single journalism piece. These function as building blocks for the broader journalistic narrative. This was particularly the case of pieces drawing on the dynamic diagrammatic mode, where these views were sequences of a larger narrative through which the reader can progress thanks to a strong interactive component realised through a diversity of techniques and devices. In addition to this, given our aim to qualitatively explore the narrative potential of networks, we have not exhaustively analysed all the narrative views elicited by networks that make up the larger narrative developed in a journalism piece. Instead we selected and limited our focus to illustrating a few narrative views that resonate with classic network concepts and properties. We do however consider important and invite researchers to study the interplay between different narrative sequences in the context of the broader narrative of a single journalism piece.

---

[24] The list is accessible here:
https://figshare.com/articles/List_of_journalism_projects_using_network_concepts_analysis_and_or_diagrams/3126523

We propose that one of the obstacles hindering the wider penetration of networks in journalistic practices is the absence of established practices for working with their narrative affordances. While over many years we have developed literacies to construe narrative meaning out of visual representations of statistical analyses, common comparisons of network visualisations with "hairballs" or "spaghetti bowls" may be considered to point to the absence of a vocabulary through which to "storify" networks. And this is not without good reason. As network techniques derive from (and are still primarily developed in) graph mathematics, it is not surprising that their analytic potential has so far received significantly more attention that their capacity to tell stories or to convey narrative meaning. Examining how networks may be read narratively is however essential given the current proliferation of network analytics and visual representations of networks in many domains of knowledge and life. And it is also crucial in order to facilitate the journalistic use of networks as meaning making instruments. We hope that the vocabulary of narrative readings as well as the protocol for the construal of narrative meaning out of networks which we developed, will serve as a first step in this direction.

Figures 1(a) and 1(b). "Connected China," *Thomson Reuters*, February 2013. (a) Part five of the visualisation guide which provides instructions for navigating the "Social Power" section of the multimedia website. (b) Network visualisation depicting the network of associations around a selected government official in China.

Figures 2(a) and 2(b). "WESD Web of Connections," *Statesman Journal*, n.d. (a) Special report title and instructions for navigating the interactive graphic. (b) Default graphic view depicting the Willamette Educational Service District and its network of connections.

Figures 3(a) and 3(b). "Top Secret America," *Washington Post*, July 19, 2010. (a) Default view of the "Explore Connections" section of the investigation with information box inviting the reader to click on an entity to see its network of connections. (b) Network of contractors and areas of work around selected government agency.

Figure 4(a) and 4(b). "The Stem Cell Wars," *New Scientist*. June 12, 2010. (a) Article title and subtitle, graphic caption and network visualisation of citations in stem cell science. (b) Information box describing how the network analysis was conducted, as well as its key findings about the influential players.

Figure 5. "Park Young-Joon at the Center of President Lee Myung-Bak's Human Resources Network,".*JoongAng Ilbo*, January 20, 2002. Article title, graphic caption and network visualisation depicting government officials central to the president's network.

Figures 6(a) and 6(b). "The Flip Investigation," *Sarasota Herald Tribune*, 2009. (a) Title and picture from article in the Flip Investigation series investigating real estate agent Craig Adams. (b) Network visualisation depicting Craig Adams at the centre of a property-flipping network in Florida.

Figure 7. "2007-2011: La Cartographie de la Blogosphère Politique," *Le Monde*. July 4, 2011. Title of journalistic piece and network graphic view of the French political blogosphere in 2011.

Figure 8. "Visualizing the Split on Toronto City Council. *Global News*, March 20, 2012. "December 2010" graphic panel depicting the group around the mayor (blue) and the opposition (red) and a few unaffiliated councillors positioned between the two clusters (purple).

Figure 9. "Mariage Gay: L'Opposition Soigne ses Amendements," *Le Monde*. January 31, 2013. Instructions for navigating the graphic and network visualisation depicting coalitions around amendments to the gay marriage law proposal in France.

Figures 10(a), 10(b) and 10(c). "2007-2011: La Cartographie de la Blogosphère Politique," *Le Monde*. July 4, 2011. (a), (b) and (c). Network mapping of the French political blogosphere at three moments in time: 2007, 2009 and 2011.

Figures 11(a), 11(b) and 11(c). "Visualizing the Split on Toronto City Council. *Global News*, March 20, 2012. (a), (b) and (c). Network visualisation of voting coalitions on the Toronto City Council at three moments in time: December 2010, January 2012 and February 2012.

Figures 12(a), 12(b) and 12(c). "How Riot Rumours Spread on Twitter," *Guardian*, December 7, 2011. (a), (b) and (c). Network diagram of clusters of tweets around the rumour "Rioters Attack a Children's Hospital in Birmingham" at three different moments in the unfolding of the rumour and its dismissal.

Figure 13. "Terrorist Tentacles Know no Boundaries," *Kansas City Star*, November 28, 2004. Tree-like network diagram depicting a global charity and its alleged terrorist ties.

Figure 14. "The Calderon Family's Connections," *Los Angeles Times*, February 21, 2014. Title, subtitle and interactive network graphic opening the journalistic piece.

Figure 15. "The Proxy Platform," Organised Crime and Corruption Reporting Project (OCCRP), November 22, 2011. Landing page of the journalistic piece depicting the interactive network graphic of actors that make up the investigated money laundering system.